\documentstyle[epsf,12pt]{article}
\begin{document}
\title       {On the effective interaction \\
                 in the light-cone QCD-Hamiltonian} 
\author{Hans-Christian Pauli \\ 
               Max-Planck-Institut f\"ur Kernphysik \\
               D-69029 Heidelberg \\ }
\date{14 July 1997}
\maketitle
\begin{abstract}
The canonical front form Hamiltonian for non-Abelian  SU(N)
gauge theory in 3+1 dimensions is mapped non-perturbatively 
on an effective  Hamiltonian which acts 
only in the Fock space of a quark and an antiquark.
Emphasis is put on  dealing with the many-body
aspects of gauge field theory, and it is shown explicitly how
the higher Fock-space amplitudes can be retrieved 
self-consistently  from solutions in the $q\bar q$-space.
The approach is based on the novel method of iterated
resolvents and on discretized light-cone quantization, 
driven to the continuum limit. It is free of the usual 
perturbative Tamm-Dancoff truncations in particle number 
and coupling constant, and respects all symmetries of the
Lagrangian including covariance and gauge invariance. 
It is emphasized that vertex  regularization as opposed to
Fock-space regularization allows an explicit 
renormalization, which yields linear confinement.
The string constant must be determined from experiment.
\end{abstract}
\vfill 
\noindent  Preprint MPIH-V26-1997 \\

\noindent   Submitted to Physical Review D\\
PACS-Index: 11.10Ef, 11.15Tk, 12.38Lg, 12.40Yx\\

\newpage
\section{Introduction} 
\label{sec:1} 

The Hamiltonian bound-state problem  in a field theory is 
notoriously difficult. Procedures like those of
Schwinger and  Dyson or  of Bethe and Salpeter are not easy 
to cope with in practice as reviewed recently in 
\cite{lsg91}.  In the Fifties, the difficulties were considered so
enormous that the Hamiltonian approach was given up 
altogether in favor of Feynman's action oriented approach. 
Its modern  successors like the lattice gauge theories
\cite{mac93,wei94} have arrived at a stage of making quantitative 
predictions \cite{nrq95}. In contrast, phenomenological  models
\cite{goi85,dkm91,neu93,eiq94}  are closer to experiment and 
have the different objective to classify the bulk of empirical data.
They leave little doubt that a heavy meson contains primarily a 
pair of constituent quarks and {\em not} an infinity of sea 
particles as suggested by the perturbative treatment of 
quantum field theory. 

How can one reconcile such models with the need
to understand hadronic structure  from a covariant theory?
Particularly, how can one reconcile the constituent quark
model (CQM) with the quantum field theory of 
chromodynamics (QCD)?
There are several reasons why the front form of Hamiltonian 
dynamics \cite{dir49} is one of the very few candidates  
(see \cite{leb80,brp91,bpp97,gla95}). 
Particularly the simple vacuum and the simple boost properties
\cite{brp91,bpp97,pab85a,pab85b} are in contrast with the
complicated vacuum and the complicated boosts in the 
conventional Hamiltonian theory (see also \cite{wil89} and 
\cite{wwh94}).
Nowadays it is known that even the front form vacuum
is not simple~\cite{hks92,vap92}, but  it is still simpler than in 
the instant form:  The problem can at least be formulated 
mathematically \cite{kal96}.

The success of discretized light-cone quantization 
\cite{pab85a,pab85b} particularly in 1+1 dimensions  has stirred 
hope that its apparent simplicities carry over to 
3+1 dimensions \cite{tab91,kpw92}. The two major obstacles
for such an endeavor are the following: (1) The Hamiltonian matrix 
increases  exponentially fast with the particle number and the 
transverse momenta in the Fock states.
Truncating the Fock space to 2 particles like
Tamm \cite{tam45}   and Dancoff \cite{dan50} invokes
perturbation theory, violates gauge and Lorentz invariance, 
and generates non-integrable singularities. 
{\it Ad hoc} procedures must be introduced to make things 
work, both in the original \cite{tam45,dan50} 
and the light-cone work \cite{kpw92}. 
(2) The unrestricted integration over the transverse momenta 
creates divergences, which are regularized by some cut-off. 
The unphysical dependence on the regularization parameters 
has to be removed by renormalization. 
This is difficult enough in a perturbative context, but serious 
challenges appear when dealing with the non-perturbative 
solutions of a Hamiltonian (see \cite{wwh94}). 

In the present work, these problems are addressed and
partially solved by adopting  systematically the point of view 
that the bound states of QCD are eigenstates of the canonical 
field-theoretical Hamiltonian in the light-cone gauge $A^+=0$. 
The zero modes emphasized in previous work 
\cite{kal96} are omitted here. 
In section~\ref{sec:2} a method for diagonalizing numerically  
a very large matrix, say with dimension $10^{21}$, by mapping  
the problem equivalently to the diagonalization of a reasonably 
small matrix, say with dimension $10^{3}$, is proposed, 
using the theory of effective interactions \cite{mof50}.
The reduction to comparatively small matrices 
is  more than a technical issue and requires one to cope
with the many-body aspects of a field theory. In the present
work,  the many-body aspects are kept as resolvents which
{\em are not expanded as perturbative series}.
They are treated in section~\ref{sec:3} by  the 
method of iterated resolvents. This novel method requires
the development of considerable formal apparatus which
was given earlier  \cite{pau96}. 
In the continuum limit the matrix equations become integral 
equations, which are similar to those solved numerically already
\cite{pam95,trp96}, but here they are derived without assuming 
that the coupling constant is small and without truncating 
the Fock space. The important issues of vertex regularization 
and renormalization are discussed in sections~\ref{sec:4} 
and~\ref{sec:5} lead ultimately to (linear) confinement.
The approach is summarized and discussed 
in section~\ref{sec:6}.  

\section{The light-cone Hamiltonian and the 
Tamm-Dancoff approach}
\label{sec:2} 

In canonical field theory the  four  commuting components of
the  energy-momentum four-vector $ P ^\nu$ are constants 
of the motion. 
In the front form of Hamiltonian dynamics \cite{dir49}
they are denoted by $ P ^\nu = (P  ^+, \vec P  _{\!\bot}, P  ^-)$, 
see for example \cite{brp91,bpp97}. 
Its spatial components 
$\vec P  _{\!\bot} =( P  ^1, P ^2)$  and $P  ^+$ 
do not depend on the interaction and are diagonal operators 
in momentum representation.
Their eigenvalues are the sums of the single particle
momentum components, {\it i.e.}
\begin {equation}
     P ^+ = \sum _{n} p^+ _{n} = {2\pi \over L } K
     \ , \quad {\rm and} \quad
     \vec P  _{\!\bot} = \sum _{n} (\vec p _{\!\bot} )_{n} 
\ . \label{eq:2.1}\end {equation}
Each particle has four-momentum 
$ p^\mu = (p^+, \vec p _{\!\bot}, p^- )$, 
sits on its mass-shell $ p^\mu p_\mu = m^2$ and therefore
has $p^- = (m^2 + \vec p ^{\,2}_{\!\bot}) / p^+ $.
Each particle state ``$q$''  is then characterized by six 
quantum numbers 
$ q = (p^+, \vec p_{\!\bot}, \lambda;c, f ) $.
The first three specify the particle's spatial momentum and 
$\lambda $ its helicity  $\uparrow$ or  $\downarrow$. 
For QED that's all there is, but for QCD a quark is specified 
further by its color and flavor, $c$ and $f$, respectively. 
Correspondingly, gluons are specified by the five quantum 
numbers $q = (p^+, \vec p_{\!\bot}, \lambda;a) $, 
with $a$ being the glue index. 
The temporal component $P ^-= 2 P_+ $ depends on the
interaction  and is a very complicated non-diagonal
operator. It propagates the system in the light-cone 
time $ x^+ = x^0 + x^3$, {\it i.e.}  
$ i { \partial \over \partial x ^+ } \vert \Psi \rangle 
   = P _+ \vert \Psi \rangle $ 
and is the proper front-form Hamiltonian \cite{dir49}.

The contraction of  $P ^\mu$  is the operator of
invariant mass-squared, {\it i.e.}
\begin {equation} 
      P ^\mu P _\mu  = P ^+ P ^- - \vec P  _{\!\bot} ^{\,2} 
      \qquad  \equiv H _{\rm LC} \equiv H
\ .\end {equation} 
It is a Lorentz scalar and  referred to
somewhat improperly but conveniently as the 
`light-cone Hamiltonian $ H _{\rm LC} $' 
\cite{brp91}, or shortly $ H $.
Its matrix elements (and eigenvalues) somewhat unusually 
carry the dimension of an invariant-mass-squared.
In this work one aims at a representation in which $H$ is diagonal, 
\begin {equation} 
      H \vert \Psi _i \rangle = E _i \vert \Psi _i \rangle
\ . \label{eq:2.2}\end {equation} 
Since the Hamiltonian does not conserve particle number,
the eigenfunctions $\vert\Psi _i \rangle$ are complicated
superpositions  of particle number projections with 
contributions like $\langle q\bar q\vert\Psi _i \rangle$ 
or $\langle q\bar q\,g\vert\Psi _i \rangle\ \cdots$. 
The eigenvalue equation (\ref{eq:2.2}) stands thus for an 
infinite set of coupled integro-differential equations which are 
even difficult to write down explicitly \cite{bpp97}. 

In order to analyze the structure of the eigenvalue equation, 
we apply a technical trick, referred to as Discretized Light Cone 
Quantization (DLCQ) \cite{pab85a,pab85b}: One imposes
periodic  boundary conditions on the Lagrangian fields.
In consequence one maps the infinite set of coupled integral 
equations onto a {\em finite set} of coupled matrix equations.
The kernel of the integral equation (\ref{eq:2.2}) becomes a 
finite matrix, which in principle can be diagonalized by 
numerical methods. 

\def\d{ T }  \def\v{ V }  \def\b{$\cdot$} \def\s{$\cdot$}   \def\f{$\cdot$}
\begin {table}[t]
\begin {center}
\caption {\label {tab:3} 
The Hamiltonian matrix $H=T+V$ for QCD and fixed value
of harmonic resolution $K=4$.~--- 
Each element of the matrix below corresponds to a block 
matrix with potentially large dimension, caused by all possible
combinations of the single particle momenta subject to fixed 
total value. The diagonal blocks are denoted by $T$. 
They are diagonal matrices, its elements being the sum of all
`kinetic energies' of the particles in the respective Fock states.
Due to the selection rules most of the block matrices 
are zero matrices, marked by a dot ($\cdot$).
The blocks marked by $V$ denote the large block matrices 
which are potentially non-zero since the vertex interaction can
get active. 
Their actual matrix elements depend on the momenta and 
helicities of the particles as tabulated elsewhere.
}\vspace{1em}
\begin {tabular}  {||cc||cc||c|ccc|cccc|ccccc||}
\hline \hline 
     &   &        & N$_p$ & 
     2 & 2 & 3 & 4 & 3 & 4 & 5 & 6 & 4 & 5 & 6 & 7 & 8 
\\ 
   K & N$_p$ & Sector & n & 
     1 & 2 & 3 & 4 & 5 & 6 & 7 & 8 & 9 &10 &11 &12 &13 
\\ \hline \hline   
    1 &  2 & $ q\bar q\, $ &  1 & 
    \d &\s &\v &\f &\b &\f &\b &\b &\b &\b &\b &\b &\b 
\\ \hline
    2 &  2 & $ g\ g\ $ &  2 & 
    \s &\d &\v &\b &\v &\f &\b &\b &\f &\b &\b &\b &\b 
\\
    2 &  3 & $ q\bar q\, \ g $ &  3 & 
    \v &\v &\d &\v &\s &\v &\f &\b &\b &\f &\b &\b &\b 
\\
    2 &  4 & $ q\bar q\, q\bar q\, $ &  4 & 
    \f &\b &\v &\d &\b &\s &\v &\f &\b &\b &\f &\b &\b 
\\ \hline
    3 &  3 & $ g \ g \ g $ &  5 & 
    \b &\v &\s &\b &\d &\v &\b &\b &\v &\f &\b &\b &\b 
\\
    3 &  4 & $ q\bar q\, \ g \ g $ &  6 & 
    \f &\f &\v &\s &\v &\d &\v &\b &\s &\v &\f &\b &\b 
\\
    3 &  5 & $ q\bar q\, q\bar q\, \ g $ &  7 & 
    \b &\b &\f &\v &\b &\v &\d &\v &\b &\s &\v &\f &\b 
\\
    3 &  6 & $ q\bar q\, q\bar q\, q\bar q\, $ &  8 & 
    \b &\b &\b &\f &\b &\b &\v &\d &\b &\b &\s &\v &\f 
\\ \hline
    4 &  4 & $ g \ g \ g \ g $ &  9 & 
    \b &\f &\b &\b &\v &\s &\b &\b &\d &\v &\b &\b &\b 
\\
    4 &  5 & $ q\bar q\, \ g \ g \ g $ & 10 & 
    \b &\b &\f &\b &\f &\v &\s &\b &\v &\d &\v &\b &\b 
\\
    4 &  6 & $ q\bar q\, q\bar q\, \ g \ g $ & 11 & 
    \b &\b &\b &\f &\b &\f &\v &\s &\b &\v &\d &\v &\b 
\\
    4 &  7 & $ q\bar q\, q\bar q\, q\bar q\, \ g $ & 12 & 
    \b &\b &\b &\b &\b &\b &\f &\v &\b &\b &\v &\d &\v 
\\
    4 &  8 & $ q\bar q\, q\bar q\, q\bar q\, q\bar q\, $ & 13 & 
    \b &\b &\b &\b &\b &\b &\b &\f &\b &\b &\b &\v &\d 
\\ \hline\hline
\end {tabular}
\end {center}
\end {table}

Why finite?
As a consequence of discretization, the Fock states are 
denumerable and orthonormal.
Since $P  ^+$ has only positive eigenvalues and  
since each particle has a lowest possible value of $p^+$, 
the number of particles in a Fock state is limited for any 
fixed value of the {\em harmonic resolution} $ K $, defined by
Eq.(\ref{eq:2.1}). One recognizes this best by way of example.
The lowest possible value $K=1$  allows only one
$q\bar q$-pairs (a single gluon cannot be in a color singlet). 
For  $K=2$, the Fock space  contains in addition pairs of gluons, 
$q\bar q$-pairs plus a gluon, and two $q\bar q$-pairs, 
as illustrated in Table~\ref{tab:3}. 
For $K=4$, one has at most 8 particles, namely 4 $q\bar
q$-pairs.  In the table all  13 Fock-space  sectors possible
for $K\leq 4$ are enumerated by $n=1,\dots,13$.
Note that the above classification does not
change  when  $K$ is increased further: one just adds more
complicated Fock-space sectors. 
Their number grows  quadratically with $K$ and 
has the value $N_K =  (K+1)(K+2)/2- 2 $. 

The single particle momenta can take all possible values 
subject to the constraints in Eq.(\ref{eq:2.1}). 
Each of the above Fock space sectors contains infinitely 
many individual Fock states, since the transverse momenta 
can take either sign. Their number can be regulated 
by some convenient ``Fock-space regularization'': 
A Fock state with $n$ particles is included only
if  the single particle four-momenta satisfy \cite{leb80}
$\left( p_1+p_2+\dots p_n\right)^2 -
      \left( m_1+m_2+\dots m_n\right)^2
      \leq \quad \Lambda_{FS}^2$. 
However, it was not realized in the past \cite{bpp97}, that 
Fock-space regularization is almost irrelevant in the 
continuum theory. One needs a better alternative, 
and we propose to work with ``vertex regularization''.
At each vertex,  a particle with four-momentum 
$p^\mu$ is scattered into two particles with respective 
four-momentum $p_1^\mu$ and $p_2^\mu$. 
One can  {\em regulate the interaction} by a condition on the 
off-shell mass ${\cal M}^2=(p_1+p_2)^2-( m_1+m_2)^2$, 
\begin{equation} 
      {\cal M}^2\leq \Lambda^2 
\,,\label{eq:i1}\end{equation} 
in terms of the regulator mass scale $\Lambda$.
This innocent step will turn out as a rather efficient tool for 
analyzing the divergences, see below.
To get more insight, the relevant momenta are
parametrized as $p^\mu=(p^+,\vec p_{\!\perp},p^-)$,  
$p_1^\mu=(zp^+,z\vec p_{\!\perp}+\vec l_{\!\perp},p_1^-)$ and
$p_2^\mu=([1-z]p^+,[1-z]\vec p_{\!\perp}-\vec l_{\!\perp},p_2^-)$. 
The off-shell mass becomes then simply
\begin{equation} 
       {\cal M}^2={1\over z(1-z)}\left( \vec l_{\!\perp}^{\ 2}
       +\left(m_1+m_2\right)^2 \left( z-\overline z \right)^2 \right) 
\,,\label{eq:i2}\end{equation} 
and the mass scale governs how much the particles can go off their 
equilibrium values ${l} _{\!\perp}=0$ and $\overline z= m_1/(m_1+m_2)$.
For $\Lambda=0$, the phase space is reduced to a point, 
and consequently the interaction is reduced to zero: 
The Hamiltonian matrix (or the integral equation) is diagonal, 
and has the spectrum of the free theory irrespective of the 
matrix dimension  governed by $\Lambda_{FS}$.

Once the Fock space is defined, one can calculate the matrix
elements of the (light-cone) Hamiltonian \cite{brp91,bpp97}. 
It is written here  as a sum of three parts, 
$      H = T + V + W $.
The kinetic energy  $T$ survives the limit of the coupling constant 
$ g $ going to zero. It is diagonal in Fock-space representation, 
its eigenvalue is the {\em free invariant mass squared}
of the particular Fock state. 
The {\em vertex interaction} $ V $ is the relativistic interaction
{\it per se}. It is linear in $g$ and changes the particle number 
by 1. Matrix elements of $V$ which
change the  particle number by 3 (as in the instant form) 
vanish kinematically:  The vacuum {\em does not fluctuate},
see for example \cite{bpp97}.
The {\em instantaneous interactions} $ W $ are 
consequences of working in the light-cone gauge,  
are {\em gauge artifacts}, and  are proportional to
$g^2$.  They either conserve particle number
(seagull interactions), or change it by 2 (fork interactions).
For the moment we will drop the instantaneous interaction $W$,
but we will restore it later. Most of the Hamiltonian matrix elements 
are then zero. Only those where the particle number
is changed by 1 are potentially non-zero, as displayed in 
Table~\ref{tab:3}. The diagonal blocks are also essentially zero
matrices, except for the kinetic energies which reside on the
diagonal. It is a long standing problem of field theory of knowing
how such a matrix can have bound states.

The key problem is that the dimension of the Hamiltonian
matrix  increases exponentially. 
Suppose, the regularization procedure allows 10 discrete
momentum states in each direction, {\it i.e.} in the one 
longitudinal and the two transverse directions of $\vec k_{\!\perp}$. 
A particle has then roughly $10^3$ degrees of freedom.
A Fock-space sector with $n$ constituent particles
has thus $10^{n-1}$ different Fock states, since they
are subject to the constraints, Eq.(\ref{eq:2.1}).
Sector 13 alone with its  8 particles has the 
dimension of  about $10^{21}$ Fock states.

The Hamiltonian method applied to gauge theory therefore 
faces a formidable matrix  diagonalization problem.
Sooner or later, the matrix dimension exceeds
imagination, and other than in 1+1 dimensions 
one has to develop new tools by matter of principle.
One would like to reduce the full problem to an effective 
interaction between a quark and an antiquark, 
\begin {equation}
     H_{\rm eff}\vert\psi_{q\bar q}\rangle 
        = E\vert\psi_{q\bar q}\rangle 
\,. \end {equation} 
The derivation of such an effective interaction, 
is the subject of this work. 

Effective interactions are a well known tool in  many-body 
physics \cite{mof50}. In field theory
the method is known as the Tamm-Dancoff-approach, since 
it was applied first to Yukawa theory for describing the 
nucleon-nucleon interaction by Tamm \cite{tam45} 
and by Dancoff \cite{dan50}. Let us review it in short.

In a field theory,  the Fock-space sectors $\vert i \rangle$ 
appear in a most  natural way, see Table~\ref{tab:3}. 
As explained above, the Hamiltonian
matrix can be understood as a matrix of block matrices,
whose rows and columns are enumerated by
$i=1,2,\dots N$ in accord with the  Fock-space sectors
like in  Table~\ref{tab:3}.
Correspondingly, one can rewrite Eq.(\ref{eq:2.2})  
as a block matrix equation:
\begin {equation} 
      \sum _{j=1} ^{N} 
      \ \langle i \vert H \vert j \rangle 
      \ \langle j \vert \Psi\rangle 
      = E\ \langle i \vert \Psi\rangle 
\, \qquad {\rm for\ all\ } i = 1,2,\dots,N 
\ .\label {eq:319}\end {equation} 
One recalls that each sector contains many individual
Fock states with different values of 
$x$, $\vec p  _{\!\bot}$ and $\lambda$.
The rows and columns of the matrix can always be split
into two parts. One speaks of  the $ P $-space  
$P = \sum _{j=1} ^n  \vert j \rangle\langle j \vert $
with $1<n<N$, and of the rest, 
the $Q$-space $Q\equiv 1-P$.
Eq.(\ref{eq:319}) can then be rewritten conveniently 
as a $2\times2$ block matrix equation
\begin {eqnarray} 
   \langle P \vert H \vert P \rangle\ \langle P \vert\Psi\rangle 
 + \langle P \vert H \vert Q \rangle\ \langle Q \vert\Psi\rangle 
 &=& E \:\langle P \vert \Psi \rangle 
 \ ,  \label{eq:321} \\   {\rm and} \qquad
   \langle Q \vert H \vert P \rangle\ \langle P \vert\Psi\rangle 
 + \langle Q \vert H \vert Q \rangle\ \langle Q \vert\Psi\rangle 
 &=& E \:\langle Q \vert \Psi \rangle 
\ . \label{eq:322}\end {eqnarray}
Rewrite the second equation as
$    \langle Q \vert E  -  H \vert Q \rangle 
    \ \langle Q \vert \Psi \rangle 
  =   \langle Q \vert H \vert P \rangle 
    \ \langle P \vert\Psi\rangle $,
and observe that the quadratic matrix 
$ \langle Q\vert E -  H \vert Q \rangle $ could be inverted 
to express the Q-space wavefunction 
$\langle Q \vert\Psi\rangle $
in terms of the $ P $-space wavefunction
$\langle P \vert\Psi\rangle$. 
But here is the problem:   
The eigenvalue $ E $ is unknown at this point. 
One therefore solves first an other problem: One introduces
{\em the starting point energy} $\omega$ as a redundant
parameter  at disposal, and 
{\em defines the $ Q $-space resolvent} or propagator
as the inverse of the block matrix 
$\langle Q \vert\omega- H \vert Q \rangle$,  thus
\begin {equation} 
         \langle Q \vert \Psi (\omega)\rangle 
  =    G _ Q (\omega) \langle Q \vert H \vert P \rangle 
         \,\langle P \vert\Psi\rangle 
\ ,\ \quad{\rm with}\quad
     G _ Q (\omega)   =  
     {1\over\langle Q \vert\omega- H \vert Q \rangle} 
\ .\label {eq:332} \end {equation} 
Inserting this into Eq.(\ref{eq:321}) defines an effective interaction
\begin{equation} 
      H _{\rm eff} (\omega) =  H +
      H \vert Q \rangle\,G_Q(\omega)\,\langle Q\vert H 
\ ,  \label{eq:340}\end{equation} 
which by construction acts only in the $P$-space
\begin{equation} 
      H _{\rm eff} (\omega ) 
      \vert P\rangle\,\langle P \vert\Psi_k(\omega)\rangle =
      E _k (\omega )\,\vert\Psi _k (\omega ) \rangle 
\ .  \label{eq:345}\end{equation} 
The effective interaction is thus well defined: It is the 
original block matrix $\langle P\vert H\vert P\rangle$ 
plus a part where the system is scattered virtually into the 
$ Q $-space, propagating there ($G_Q$) by impact of the 
true interaction, and scattered back into the $ P $-space. 
Every value of $\omega$ defines a different Hamiltonian 
and a different spectrum. 
Varying $\omega$ one generates a set of  
{\em energy functions} $ E _k(\omega) $. 
Whenever one finds a solution to the {\em fix point equation} 
\begin {equation}
E _k (\omega ) = \omega 
, \label {eq:350} \end {equation}
one has found one of the true eigenvalues and
eigenfunctions of $H$,  by construction. 

One should emphasize that one can find all eigenvalues 
of the full Hamiltonian $H$,  
irrespective of how small one chooses the $ P $-space. 
Explicit examples for that can be found in \cite{pau96}. 
It looks as if one has mapped a difficult problem, the 
diagonalization of a huge matrix onto a simpler problem,  
the diagonalization of a much smaller matrix. 
But this is only true in a restricted sense. 
One has to invert a matrix. The numerical inversion of a
matrix takes  about the same effort as its diagonalization. 
In addition,  one has to vary $\omega$ and solve the fix point 
equation (\ref{eq:350}).  The numerical work is thus larger 
than the direct diagonalization. 

The advantage of working with an effective interaction is 
of an analytical nature because the resolvents can be
approximated systematically.  The two resolvents
\begin {equation}
     G _Q (\omega) =  {1\over \langle Q \vert 
           \omega - T - U  \vert Q \rangle} 
     \quad{\rm and}\quad 
     G _0  (\omega) =  {1\over \langle Q \vert 
           \omega - T \vert Q \rangle} 
\ , \label {eq:352} \end {equation} 
defined once with and once without the non-diagonal 
interaction $U$, are identically related by 
$G_Q(\omega)=G_0(\omega)+G_0(\omega)\,U\,G_Q(\omega)$,
or by the infinite series of perturbation theory.
The point is, of course, that the kinetic energy $T$ is
a diagonal operator which can be trivially inverted to get 
the unperturbed resolvent $G_0 (\omega)$. 
In practice \cite{kpw92,tam45,dan50,trp96}, one has never 
gone beyond the first non-trivial order of perturbation theory 
for formulating the effective interaction, namely 
$H _{\rm eff} (\omega) =  T_{q\bar q} + W+V\vert  
  q\bar q\,g\rangle\,G_0(\omega)\,\langle q\bar q\,g\vert V$.
This, of course, is highly unsatisfactory.

The identification of the
$q\bar q$-sector with the $P$-space and the related 
effective interaction appears as the most natural thing to do. 
But in practice, the Tamm-Dancoff-approach (TDA)
has two serious defects: (1) The approach
is technically useful only if one truncates the perturbation 
series to its very first term. This destroys
Lorentz and gauge invariance. (2) If one identifies 
$\omega$ with the eigenvalue as one should, the effective 
interaction develops a non-integrable  singularity, 
both in the original instant form \cite{tam45,dan50} and 
in the front form \cite{kpw92,trp96}.
In order to force the approach to work, one has to replace 
the eigenvalue $\omega$ by the kinetic energy in the 
$P$-space $T_{q\bar q}$ \cite{kpw92,tam45,dan50}.

\section {The effective interaction and the wavefunction}
\label {sec:3}

In gauge theory the Fock space sectors appear in the most 
natural way, and in DLCQ the number of  Fock-space 
sectors is finite from the outset as shown in Table~\ref{tab:3}. 
The block matrix is sparse due to the nature of the underlying 
Hamiltonian, most of the block matrices are zero matrices.
One should make use of that!
The Tamm-Dancoff approach, particularly the step from 
Eqs.(\ref{eq:321})  and (\ref{eq:322})  to Eq.(\ref{eq:345}),
can be interpreted as the reduction  of a block matrix
dimension  from 2 to 1. But there is no need for identifying the 
$P$-space with the lowest sector. Nothing prevents us from
choosing the last sector as the $Q$-space: 
The same steps as above reduce then the block matrix dimension 
from $N$ to $N-1$. The effective interaction acts in the now 
smaller space. This procedure can be iterated. The price to pay
is that one has to deal with `resolvents of resolvents', 
or with iterated resolvents. 
Ultimately, one arrives at block matrix dimension 1 where the 
procedure stops: The effective interaction in the Fock-space
sector with only one quark and one antiquark is defined
unambiguously.
In order to make progress one needs to readjust notation 
\cite{pau96}. Suppose, in the course of this reduction, one has 
arrived at block matrix dimension $n$, with $1\leq n\leq N$.  
Denote the corresponding effective interaction  $H_n (\omega)$. 
Since one starts from the full  Hamiltonian in the last 
sector $N$, one has to convene that $H_{N}\equiv H$. 
The eigenvalue problem corresponding to  
Eq.(\ref{eq:345}) reads then
\begin {equation} 
   \sum _{j=1} ^{n} \langle i \vert H _n (\omega)\vert j \rangle 
                      \langle j \vert\Psi  (\omega)\rangle 
   =  E (\omega)\ \langle i \vert\Psi (\omega)\rangle 
\ , \ \quad{\rm for}\ i=1,2,\dots,n
\ . \label {eq:406} \end {equation}
Observe that $i$ and $j$ refer here to sector numbers.  
Now,  in analogy to Eq.(\ref{eq:332}), define 
\begin {eqnarray} 
   \langle n \vert \Psi (\omega)\rangle   
   =  G _ n (\omega) 
   \sum _{j=1} ^{n-1} \langle n \vert H _n (\omega)\vert j \rangle 
   \ \langle j \vert \Psi (\omega) \rangle 
\ ,\quad{\rm with}\quad
    G _ n (\omega)   
    =   {1\over \langle n \vert\omega- H_n (\omega)\vert n \rangle} 
\ . \label {eq:410} \end {eqnarray}
Note that this definition requires only the inversion of the 
sector Hamiltonian $H_n(\omega)$, but not of the total Hamiltonian.
The effective interaction 
in the  ($n -1$)-space becomes then 
\begin {equation}  
       H _{n -1} (\omega) =  H _n (\omega)
  +  H _n(\omega) G _ n  (\omega) H _n (\omega)
\ .\label {eq:414} \end {equation}
This holds for every block matrix  element 
$\langle i \vert H _{n-1}(\omega)\vert j \rangle$.  
To get the corresponding eigenvalue equation one
substitutes $n$ by $n-1$ everywhere in Eq.(\ref{eq:406}). 
Everything proceeds like in section~\ref{sec:2}, 
including the fixed point equation  $ E  (\omega ) = \omega $.
But one has achieved much more: Eq.(\ref{eq:414}) is a 
{\em recursion relation} which holds for all $1<n<N$!

\begin{figure} [t]
\begin{minipage}[t]{80mm} 
\makebox[0mm]{}
\epsfxsize=80mm\epsfbox{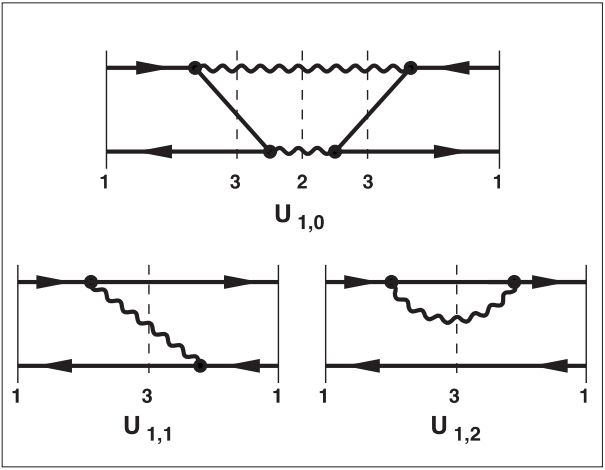}
\vfill \end{minipage}
\hfill
\begin{minipage}[t]{80mm} \makebox[0mm]{}
\makebox[0mm]{}
\caption{\label{fig:6_1} \em
The three graphs of the effective interaction in the 
$q\bar q$-space.
The lower two graphs correspond to the chain $U=VG_3V$,
the upper corresponds to $U_a=VG_3V G_2V G_3V$. 
Propagator boxes are represented by vertical dashed lines, 
with the subscript `$n$'  referring to the sector numbers. 
} \vfill \end{minipage}
\end{figure}

The method of iterated resolvents was applied in \cite{pau96}
to the paradigmatic case of a $4\times 4$-matrix, 
including explicit numerical examples. 
Applying the method to the block matrix structure of QCD, 
as displayed in Table~\ref{tab:3}, is particularly easy and 
transparent.
By definition, the last sector contains only the diagonal kinetic 
energy, thus  $H_{13}=T_{13}$. Its resolvent is calculated trivially. 
Then $H_{12}$ can be constructed 
unambiguously, followed by $H_{11}$, and so on, until one 
arrives at sector 1. Grouping them in a different 
order,  one finds for the sectors with one $q\bar q$-pair: 
\begin {eqnarray} 
     H_{q\bar q} = \qquad     H _1 
&=&  T_1 +\phantom{[}V G _3 V 
   + V G _3 V  G _2 V G _3 V \phantom{]}
\ , \label{eq:610}\\  
     H_{q\bar q\,g} = \qquad     H _3 
&=&  T_3 + \phantom{[}V G _6 V + V G _6 V  G _5 V G _6 V 
    \phantom{]} + V G _4 V 
\ , \label{eq:620}  
\\  
     H_{q\bar q\,gg} = \qquad  H _6 
&=&   T_6 + \phantom{[}V G _ {10} V+ V G _{10} V  G _9 V G _ {10} V 
    \phantom{]} + V G _7 V 
\ , 
\\
     H_{q\bar q\,ggg} = \qquad  H _{10} 
&=&  T_{10} + \left[ V G _{15} V + V G _{15} V  G _{14} V G _{15}V\right]
      +V G _{11} V  
\ . \label {eq:622} \end {eqnarray}
The quark-gluon content of  the respective sectors is added here 
for an easier identification. The square bracket in the last expression
indicates how the effective interaction looks for a higher $K$; 
for $K=4$ the content of the square bracket has to be set to zero. 
Correspondingly, one obtains  for  the sectors with two 
$q\bar q $-pairs: 
\begin {eqnarray} 
     H_{q\bar q q\bar q} = \qquad     H _4 
&=&  T_4 + \phantom{[}V G _7 V 
   + V G _7 V  G _6 V G _7 V  \phantom{]}
\ , 
\\  
     H_{q\bar q q\bar q\, g} = \qquad     H _7 
&=&   T_7 + \phantom{[}V G _ {11} V+ V G _{11} V  G _{10} V G _ {11} V 
            \phantom{]}     + V G _8 V 
\ , 
\\
     H_{q\bar q q\bar q\, gg} = \qquad     H _{11} 
&=&  T_{11} + \left[ V G _{16} V + V G _{16} V  G _{15} V G _{16}V\right]
                 + V G _{12} V 
\ , 
\end {eqnarray}
and for those with with three $q\bar q $-pairs
\begin {eqnarray} 
     H_{q\bar q q\bar q q\bar q } = \qquad     H _8 
&=&  T_{8} + V G _{12} V 
   + \phantom{[}V G _{12} V  G _{11} V G _{12}V\phantom{]}
\ , 
\\
     H_{q\bar q q\bar q q\bar q\, g} = \qquad     H _{12}
&=&  T_{12} + \left[ V G _{17} V + V G _{17} V  G _{16} V G _{12}V\right]
                 + V G _{13} V 
\ . 
\end {eqnarray}
In either case, their structure is different from the 
one in the pure glue sectors which is
\begin {eqnarray} 
     H_{gg} = \qquad     H _2 
&=&  T_2 + V G _3 V + \phantom{[}V G _5 V \phantom{]}
\ ,\label{eq:643} 
\\  
     H_{gg\, g} = \qquad     H _5 
&=&   T_5 + V G _ {6} V + \phantom{[}V G _{9} V \phantom{]}
\ , 
\\
     H_{gg\, gg} = \qquad     H _9 
&=&  T_{9} + V G _{10} V + \left[ V G _{14} V \right]
\ . \label{eq:645}\end {eqnarray}
Having arrived at this point, one can restore first the limit 
$K\longrightarrow\infty$,
simply by suppressing the square brackets in the above relations.
Second, the instantaneous interaction $W$ omitted thus far can 
be restored {\it ex post}.
Due to the peculiar structure of the gauge theory Hamiltonian displayed
in Table~\ref{tab:3}, the vertex interaction appears only in even pairs, 
typically in the combination  $VGV$. It is plausible that  by
simply substituting $VGV\rightarrow W+VGV$ restores the full
Hamiltonian $H=T+V+W$, as was checked explicitly in \cite{pau96}.

The full effective Hamiltonian $ H_{\rm eff} = H_{q\bar q}$ is 
given by Eq.(\ref{eq:610}) and displayed diagrammatically
in Fig.~\ref{fig:6_1}. What does it mean physically? 
The first term is the kinetic energy ($T_1$) in the 
$q\bar q$-space.   
In the second term ($V G _3 V$), the vertex interaction $V$ 
creates a gluon and scatters the system from the  
$q\bar q$- into the $q\bar q\,g$-space. 
As indicated in Fig.~\ref{fig:6_1}
by the vertical line with the subscript `3', the three
particles propagate there in `3'-space under impact of the full
interaction before the gluon is  absorbed. 
The gluon can be absorbed either by the antiquark or by the 
quark, represented by the two respective graphs $U_{1,1}$ 
and $U_{1,2}$ in the figure. 
The third term ($V G _3 V G _2 V G _3 V $), finally, 
in Eq.(\ref{eq:610}) describes the 
virtual  annihilation of the $q\bar q$-pair into two gluons, and
is represented by the graph $U_{1,0}$ in Fig.~\ref{fig:6_1}. 
It generates an interaction between different quark flavors.
As a net result the interaction scatters a quark with helicity
$\lambda_q$ and four-momentum 
$p = (xP^+, x\vec P_{\!\perp} + \vec k_{\!\perp}, p^-)$
into a state with $\lambda_q^\prime$ and four-momentum 
${p^\prime} = (x^\prime P^+, x^\prime\vec P_{\!\perp} 
+ \vec k_{\!\perp}^\prime, {p^\prime}^-)$, 
and correspondingly the antiquark.

In the continuum limit, the resolvents are replaced by 
propagators and the eigenvalue problem 
$H _{\rm eff}\vert \psi\rangle=M^2\vert \psi\rangle$ 
becomes again an integral equation but one with a
more transparent structure:
\begin{eqnarray} 
    M_b^2\langle x,\vec k_{\!\perp}; \lambda_{q},
    \lambda_{\bar q}  \vert \psi _b\rangle =
    \left[ {\overline m^2_{q} + \vec k_{\!\perp}^2 \over x } +
    {\overline m^2_{\bar  q} + \vec k_{\!\perp}^2 \over 1-x } \right]
    \langle x,\vec k_{\!\perp}; \lambda_{q},
    \lambda_{\bar q}  \vert \psi _b\rangle &&
\nonumber\\
    +\sum _{ \lambda_q^\prime,\lambda_{\bar q}^\prime}
    \!\int_D\!dx^\prime d^2 \vec k_{\!\perp}^\prime\,
    \langle x,\vec k_{\!\perp}; \lambda_{q}, \lambda_{\bar q}
    \vert U+U_a\vert x^\prime,\vec k_{\!\perp}^\prime; 
    \lambda_{q}^\prime, \lambda_{\bar q}^\prime\rangle\,
    \langle x^\prime,\vec k_{\!\perp}^\prime; 
    \lambda_{q}^\prime,\lambda_{\bar q}^\prime  
    \vert \psi _b\rangle &&
\ .\label{eq:445}\end {eqnarray}
The effective quark mass $\overline  m$ includes the 
impact of diagram $U_{1,2}$.  
The domain $D$ restricts integration in line with regularization 
(see below). The effective interaction $U$ includes all fine 
and hyperfine interactions.
The flavor-changing annihilation interaction 
$U_a=U_{1,0}$ is probably of less importance in a first assault
and will be disregarded in the sequel.
The eigenvalues refer to the invariant mass $M_b$ of a physical 
state and the corresponding wavefunction 
$      \langle x,\vec k_{\!\perp}; \lambda_{q},
        \lambda_{\bar q}  \vert \psi _b\rangle$ 
gives the probability amplitudes for finding in that state
a flavored quark with momentum fraction $x$, intrinsic transverse 
momentum $\vec k_{\!\perp}$ and helicity $\lambda_{q}$. 
Both are boost-invariant quantities. Color plays a subordinate 
role because the Fock states can be made color-neutral.

The eigenfunctions $\psi_b$ represent the normalized projections 
of the full eigenfunction $\vert\Psi\rangle$ onto the Fock states
$ \vert q;\bar q\rangle 
   = b^\dagger _q d^\dagger_{\bar q} \vert vac \rangle $. 
Their knowledge is  sufficient to retrieve all desired Fock-space  
components of the total wavefunction $\vert\Psi\rangle$. 
Let us work that out explicitly by asking for the probability 
amplitude to find a  $\vert gg\rangle$- or a  
$\vert q\bar q\,g\rangle$-state
in an eigenstate $\vert\Psi\rangle$ of the full Hamiltonian. 
The key is the upwards recursion relation Eq.(\ref{eq:410}).
Obviously, one can express the higher Fock-space components 
$\langle n\vert\Psi\rangle$ as functionals of $\psi_{q\bar q}$ 
by a finite series of quadratures, {\it i.e.} of matrix multiplications 
or of momentum-space integrations.  
One needs not solve another eigenvalue problem.  
The first two equations of the recursive set in Eq.(\ref{eq:410}) are
\begin {eqnarray} 
        \langle 2 \vert\Psi\rangle &=& 
        G _ 2 \langle2\vert H_2\vert1\rangle 
        \langle 1 \vert \Psi  \rangle 
\,,\label{eq:448}\\  {\rm and}\quad
        \langle 3 \vert\Psi\rangle  &=& 
        G _ 3 \langle 3 \vert H _3 \vert 1 \rangle 
        \langle 1 \vert \Psi  \rangle +
        G _ 3 \langle 3 \vert H _3 \vert 2 \rangle 
        \langle 2 \vert \Psi  \rangle 
\,.\end{eqnarray}
The sector Hamiltonians $H _n$ have to be substituted from
Eqs.(\ref{eq:620}) and (\ref{eq:643}).
In taking block matrix elements of them, the formal expressions
are simplified considerably since many of the Hamiltonian blocks 
in Table~\ref{tab:3} are zero.  
One thus gets simply
$\langle2\vert H _2\vert1\rangle=\langle2\vert VG_3V\vert1\rangle$
and therefore
$\langle2\vert\Psi\rangle=G_2VG _3V \langle1\vert\Psi \rangle$.
Substituting this into Eq.(\ref{eq:448}) gives 
$\langle 3 \vert \Psi \rangle  = G _ 3 V\langle 1 \vert\Psi \rangle +
    G _ 3 VG _ 2 VG _3 V \langle1\vert\Psi\rangle$. 
These findings can be summarized more readable as
\begin {eqnarray} 
    \vert \psi_{gg} \rangle  &=& 
    G _ {gg} VG _{q\bar q\,g}V\:\vert\psi_{q\bar q}\rangle  , 
\\  {\rm and } \qquad
    \vert\psi_{q\bar q\,g}\rangle  &=& 
    G _ {q\bar q\,g} V\:\vert\psi _{q\bar q}\rangle +
    G _ {q\bar q\,g} VG _ {gg} VG _{q\bar q\,g} V 
    \:\vert\psi_{q\bar q}\rangle 
\ .\end{eqnarray}
Correspondingly, one can write down the higher Fock-space
wavefunctions as functionals of only $\vert \psi_{q\bar q} \rangle$.
Of course, one has to readjust the overall normalization of the 
wavefunction, and this depends on how many Fock spaces one
decides to include. 
Finally, it should be emphasized again that the finite number of 
terms contrasts strongly to the infinite number of terms in 
perturbative series, because the iterated resolvents re-sum the 
series to all orders in closed form. The above results are all exact. 

Aiming for the higher Fock space amplitudes is not an esoteric
problem. One needs them for calculating the structure functions,
for example. They are the probability  amplitudes for finding
in a hadron a quark with longitudinal momentum fraction 
$x$, irrespective of its transverse momentum, and to get
them, one must sum over all Fock-space amplitudes
$f(x) = \sum_n \langle\overline\Psi_n\vert b^\dagger (x) b(x)
\vert\Psi_n\rangle$, see for example \cite{bpp97}.

\begin{figure} [t]
\begin{minipage}[t]{80mm} 
\makebox[0mm]{}
\epsfxsize=80mm\epsfbox{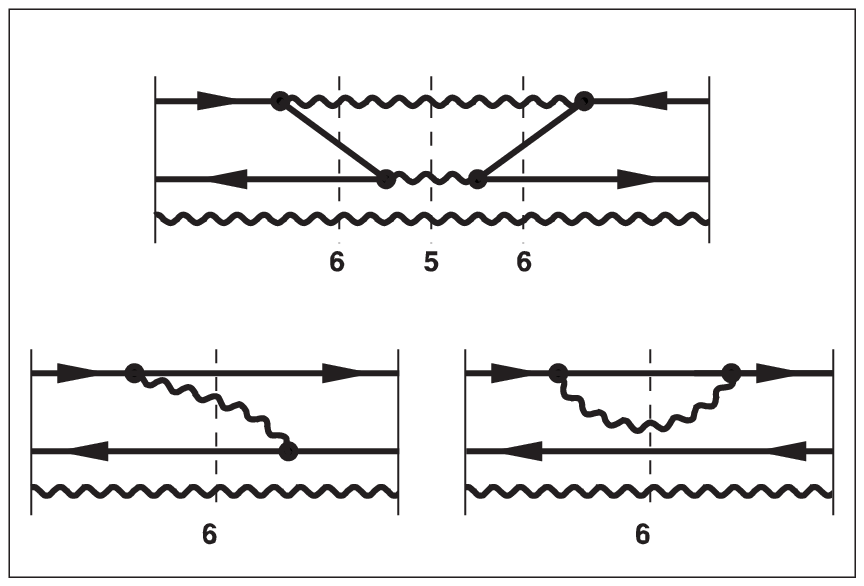}
\caption{\label{fig:6_2} \sl
    The three graphs of the  spectator interaction 
    in the $q\bar q\,g$-space. 
    Note the role of the gluon  as a spec\-tator.
}\vfill\caption{\label{fig:6_3} \sl
    Some six graphs of the  participant interaction 
    in the $q\bar q\,g$-space.
}\vfill\end{minipage}
\hfill
\begin{minipage}[t]{80mm} 
\makebox[0mm]{}
\epsfxsize=80mm\epsfbox{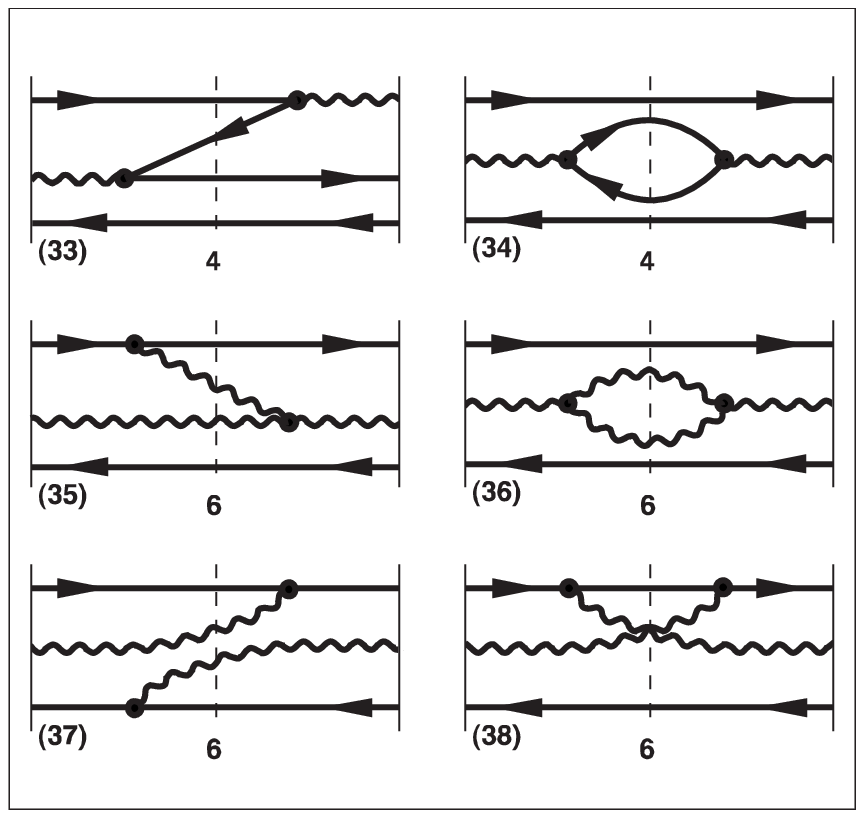}
\vfill \end{minipage}
\end{figure}

Let us discuss in some greater detail the structure of the
sector Hamiltonians, particularly the effective interaction
in the ${q\bar q\,g}$-space as given by Eq.(\ref{eq:620}).
The corresponding graphs are displayed 
diagrammatically in Figs.~\ref{fig:6_2}  and \ref{fig:6_3}. 
Those in Fig.~\ref{fig:6_2} differ from those in 
Fig.~\ref{fig:6_1} only by an additional gluon.
The gluon does not change quantum numbers under impact
of the interaction and acts like a spectator.  Therefore, 
the graphs in Fig.~\ref{fig:6_2} will be referred to as the  
`spectator interaction' $\overline U _3$. 
In the graphs of Fig.~\ref{fig:6_3}, the gluons are scattered
by the interaction, and correspondingly these graphs will be 
referred to as the `participant interaction' $\widetilde U _3$.  
The analogous separation into spectators and participants  
can be made in all quark-pair-glue sector Hamiltonians:
\begin {equation}  
     H _n =  T  _n + \overline U _n + \widetilde U _n
\ , \quad{\rm for}\quad  n= 3,6,10,15,\dots
\ .  \label {eq:626} \end {equation} 
The Hamiltonian is additive in spectators and participants. 
The spectator interaction $\overline U _n$ therefore can be 
associated with its own resolvent $\overline G _n$,
\begin{equation} 
     \overline G _n  =   {1\over \omega-T_n-\overline U_n }
\,, \qquad \biggl( {\rm while\ } G _n\equiv{1\over \omega-H_n} 
     ={1\over \omega-T_n-\overline U_n-\widetilde U_n}\ \biggr)
\,, \label{eq:641}\end{equation}  
with an exact relation to the full resolvents $G_n$, {\it i.e.} 
$G_n=\overline G_n+\overline G_n\,\widetilde U_n\,G_n$.         
Equivalently, one can write it as an infinite series
$ G _n  = \overline G _n
     + \overline G _n \widetilde U _n\, \overline G _n 	
     + \overline G _n \widetilde U _n\, \overline G _n 
                                     \widetilde U _n\, \overline G _n   + \dots $, 
as usual. The difference to the usual perturbative 
series in Sec.~\ref{sec:2} is, that there the `unperturbed 
propagator' $ G _0 (\omega) $ refers to the system without 
interactions while here the `unperturbed propagators' 
$\overline G _n$ contain the
interaction in the well defined form of $\overline U_n$.
One therefore deals here with a `perturbation theory in medium'.
Note that the present series implies also different physics: 
The system is not scattered into other sectors but 
stays in sector $n$. This is reflected in the operator identity,
which is obtained straightforwardly from
Eqs.(\ref{eq:626})  and (\ref{eq:641}):
\begin  {equation} 
  (1-\widetilde U _n \overline G _n )(\omega - \overline H _n ) =
  \left( 1-\widetilde U _n {1\over \omega - \overline H _n}
  \right) (\omega - \overline H _n ) = \omega - H _n 
\ .\label{eq:6530} \end {equation} 
Taking the inverse gives $G _n = \overline G _n R_n^2$, thus
\begin  {equation} 
      R_n^2 =  {1\over 1 - \overline G _n \widetilde U _n}
\ . \label{eq:6520}\end {equation} 
Since $R_n$ and $\overline G _n$ commute, one ends up with 
$G _n = R_n\,\overline G _n \,R_n$.
In all of the above effective interactions the square matrices 
$R_n$ are sandwiched between a quark-pair-glue resolvent 
$\overline G$ and the vertex $V$,
\begin  {equation} 
     V\,G _n \, V = V\,R_n\,\overline G _n \,R_n\, V 
     = \overline V \,\overline G_n\,\overline V      
\,,\label {eq:652} \end {equation} 
with $\overline V = R_n V$ being an effective vertex. 
The index $n$ needs not to be carried on. One can thus
rewrite systematically Eqs.(\ref{eq:610})-(\ref{eq:622}). 
Working backwards yields
\begin {eqnarray} 
     \overline H _6 
&=&   T_6 + \overline V \,\overline G _ {10} \overline V
                     + \overline V \,\overline G _{10} \overline V  
              G _9 \overline V \,\overline G _ {10} \overline V 
\ , \label {eq:662} \\
     \overline H _3 
&=&  T_3 + \overline V \,\overline G _6 \overline V 
                    + \overline V \,\overline G _6 \overline V  
              G _5\overline V \,\overline G _6 \overline V 
\ , \label {eq:663} \\      
       H _1= \overline H _1
&=&  T_1 + \overline V \,\overline G _3 \overline V 
                    + \overline V \,\overline G _3 \overline V  
             G _2 \overline V \,\overline G _3 \overline V 
\ . \label {eq:664} \end {eqnarray}
Instead of being similar, the quark-pair-glue sector
Hamiltonians  $\overline H _n = T_n + \overline U _n$ 
are now all identical. Note that, in the strict sense, they are 
equal  only in the continuum limit. For any finite $K$, there is
a last sector which is creating edge effects. 
The resolvents $G _2$, $G _5$ and $G _9$ carry no bar. 
In the pure glue sectors, the distinction between participants
and spectators makes no sense.

\section{Analysis of propagators and vertex functions}
\label{sec:4}

Thus far the approach is formally exact. In the sequel,  
this rigor is given up in favor of transparency. For calculating 
the effective interaction $U$ one needs the resolvents 
$\overline G_3$. For calculating $\overline G_3$, one needs 
$\overline G_4$ and $\overline G_6$, and so one.
Does this ever come to an end? A careful analysis shows
that there is a simple relation between them since all 
$\overline H_n(\omega)$  are  all {\it bona fide} Hamiltonians. 
Suppose we have solved the eigenvalue problem in the 
$q\bar q$-space, 
\begin  {equation}
    \sum _{q^\prime,\bar q^\prime}
    \langle q;\bar q\vert H_{q\bar q} (\omega)
    \vert q^\prime;\bar q^\prime\rangle\,
    \langle q^\prime;\bar q^\prime\vert
    \psi_b(\omega) \rangle =  M_{b} ^2(\omega) 
    \langle q;\bar q\vert\psi_b(\omega) \rangle
\,.\label{eq:6.60}\end {equation} 
The eigenvalues are enumerated by
$b=1,2,\dots\infty$. The  eigenfunctions 
$\langle q;\bar q\vert\psi_b(\omega) \rangle$ are a complete set.  
Despite working in the continuum limit, we continue to use 
summation symbols for the sake of a more compact notation. 
Suppose further that an $\omega$ was found 
which has the same value as the lowest eigenvalue $M^2=M^2_1$. 
The substitution $\omega=M^2$ will hence forward be done
without explicitly mentioning.
Next, ask for the eigenvalues and eigenfunctions
in the $q\bar q\,g$-space: 
\begin  {equation}
    \sum _{q^\prime,\bar q^\prime,g^\prime}
    \langle q;\bar q;g\vert
    \overline H_{3} \vert q^\prime;\bar q^\prime;g^\prime\rangle
    \langle q^\prime;\bar q^\prime;g^\prime\vert\psi_{b,s}\rangle
    = M_{b,s} ^2
    \langle q;\bar q;g\vert\psi_{b,s}\rangle
\,.\end {equation} 
It is important to note that one needs not to perform 
another calculation, becuase the result is already known.
By construction, the gluon is a free particle which moves
relative to the meson subject to momentum conservation.
The eigenfunction is a product state 
$\vert \psi_{b,s}\rangle=\vert \psi_{b}\rangle
   \otimes \vert \varphi_{s}\rangle$. 
Parameterizing the  gluon's four-momentum as 
\begin{equation}
   q_g^\mu=(yP^+,y\vec P_{\!\perp}+\vec q_{\!\perp},p_g^-)
\label{eq:6.690}\end{equation}
implies that the gluon moves with momentum fraction $y$ and 
transverse momentum $\vec q_{\!\perp}$.
It also implies that the meson  moves with momentum fraction 
$1-y$ and transverse momentum $-\vec q_{\!\perp}$.
Since the gluons are massless, the eigenvalues are therefore
\begin  {equation}
    M_{b,s} ^2
    = {M_b^2 + \vec q_{\!\perp}^{\,2} \over (1-y)} 
    + {\vec q_{\!\perp}^{\,2} \over y} 
\,.\end {equation}
Knowing the eigenvalues and eigenfunctions, one can
calculate the exact resolvent. A few identical rewritings give
\begin{eqnarray}
      \langle q;\bar q;g \vert \overline G _3
      \vert q ^\prime;\bar q^\prime;g ^\prime \rangle &=&
      \overline G _3 (q;\bar q;g)\left[
      \langle q;\bar q;g            \vert q ^\prime;\bar q^\prime;g^\prime\rangle -
      \langle q;\bar q;g\vert A\vert q ^\prime;\bar q^\prime;g^\prime\rangle 
      \right]\,,\qquad{\rm with}
\label{eq:6.69}\\
    A&=& \sum _{b,s} 
    \vert\psi_{b,s}\rangle\,{y(M_b^2-M^2)
    \over  Q^2+y(M_b^2-M^2)}\, 
    \langle \psi_{b,s} \vert 
\,,\\
    \overline G _3 (q;\bar q;g) &=& -{y\over Q^2}
\ , \quad{\rm and}\quad
    Q^2 = (y^2 M ^2 +\vec q_{\!\perp}^{\,2}){1\over 1-y}
\,.\label{eq:6.68}\end{eqnarray}
Since the operator $A$ by definition 
cannot become a Dirac-$\delta$ function 
$\langle q;\bar q;g\vert q^\prime;\bar q^\prime;g^\prime\rangle$,
one can drop $A$, as an approximation. 
The same type of approximation parametrizes the two gluon 
propagator $\overline G_2 (g_1;g_2)= (M^2 - M^2_{gb})^{-1}$
in terms of the glue ball mass $M^2_{gb}$.

Let us digress for a short discussion. 
One should note first that the above approximation 
can be {\em controlled e posteriori}:  
Once the eigenfunctions in the $q\bar q$-sector are available
numerically, one can check whether  the exact definition of the 
resolvent Eq.(\ref{eq:6.69}) is peaked like a $\delta$-function, 
with a residue predicted by Eq.(\ref{eq:6.68}). 
Second, the above notation is suggestive  
for  $Q^2$ being related to the single-particle four-momentum
transfer along the quark line in graph $U_{1,1}$ in 
Figure~\ref{fig:6_1}.  In fact one gets 
$(p-p^\prime)^2= -[y^2(2m)^2 +\vec q_{\!\perp}^{\,2}]$, 
for sufficiently small $y$. 
The momentum transfer is the same as in Eq.(\ref{eq:6.68})
if one replaces the `current mass' $m$ by the `constituent mass' 
$M/2$. Apart from that the exact resolvent agrees with the 
perturbative propagator:  In the solution, the particles 
propagate  like free particles.
Third, and most importantly, instead of having resolvents of 
resolvents, the hierarchy of iterated resolvents is broken. 
{\it Ex post}, this justifies the {\it ad-hoc} procedures in the 
Tamm-Dancoff approach, 
both in the original work \cite{tam45,dan50} 
and in its light-cone adaption \cite{kpw92,trp96}. 
Last but not least,
one should emphasize that the complicated operator
$A$ in Eq.(\ref{eq:6.69}) can perhaps be omitted when dealing 
with the low energy part of the spectrum. 
This becomes inadequate if not false at sufficiently high excitations, 
at sufficiently  large $\omega$, 
where one might be in the  region of `overlapping resonances'
\cite{vwz85}. This requires different techniques, possibly those 
based on random matrix models \cite{ver94,wsw96}. 
It is here where the {\em concept of a  temperature} will possibly 
make its way into the Hamiltonian bound-state problem.

\begin{figure} [t]
\begin{minipage}[t]{65mm} 
\makebox[0mm]{}
\begin{center}
\epsfxsize=60mm\epsfbox{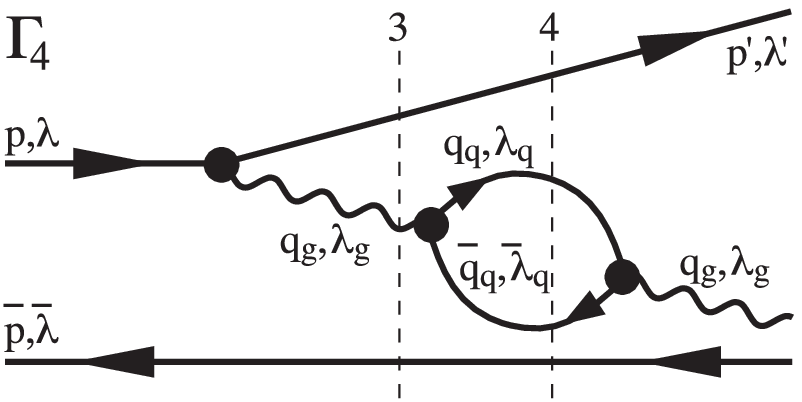}
\end{center}
\caption{\label{fig:g_1} 
The $q\bar q$ vacuum polarization graph.
}\vfill\end{minipage}
\hfill
\begin{minipage}[t]{65mm} 
\makebox[0mm]{}
\begin{center}
\epsfxsize=60mm\epsfbox{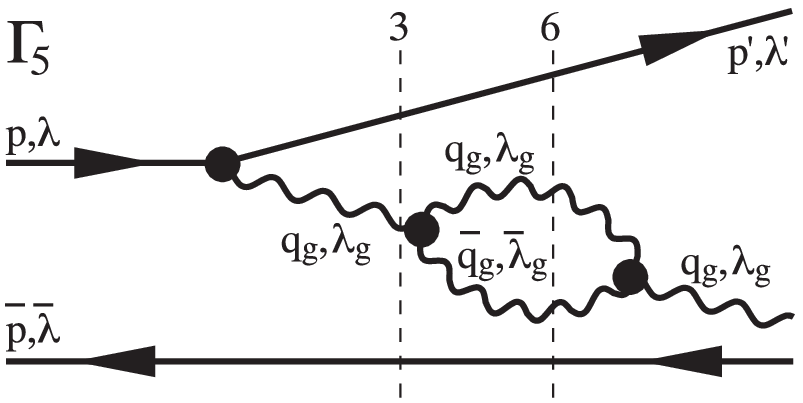}
\end{center}
\caption{\label{fig:g_2} 
The $gg$ vacuum polarization graph.
}\vfill \end{minipage}
\end{figure}

Next, how can one approximate the vertex functions $R_n$?
With Eq.(\ref{eq:6530}) one has
\begin  {equation}
       R_n  =   1
  +  {1\over 2} \overline G _n \widetilde U _n
  +  {3\over 8} \overline G _n \widetilde U _n   \,
                           \overline G _n \widetilde U _n + \dots  
\,.  \end{equation} 
For $n=3$, this reads 
$R_3  =  (1-\overline G_3 [ VR_6\overline G _6 R_6 V 
     +VR_4\overline G _4 R_4 V ]) ^ {-{1\over2} }$.
Expanding up to the first non-trivial term in the expansion gives
\begin{equation}
    R_3\simeq 1+ {1\over2} \left(\overline G_3V\overline G_4V+
    \overline G _3 V\overline G _6 V \right) = 1 + {1\over2}  \left(
    \Gamma_1+\Gamma_2+\Gamma_3+\Gamma_4+\Gamma_5
\right)\ .\end{equation}
For being consistent one must set $R_4=R_6=1$. 
Unfortunately, this was not realized 
in the previous work \cite{pau96}.
The vertex function $R_3$ identifies itself as the conventional
vertex function. 
The altogether 5 diagrams of light-cone QCD had been 
calculated before by Thorn \cite{tho79} and by Perry \cite{phz93}. 
Because they are not available in the form we need them in 
the sequel, we have recalculated them and present them 
elsewhere \cite{rap97}. 
In the nomenclature of \cite{phz93}, the mass term  
$\Gamma_1$ cancels almost completely with the two vertex 
corrections $\Gamma_2$ and $\Gamma_3$. In the sequel
we consider explicitly only the two vacuum polarization diagrams 
$\Gamma_4$ and $\Gamma_5$, as given in Figs.~\ref{fig:g_1} 
and \ref{fig:g_2}. Parameterizing the particle momenta as 
\begin  {eqnarray}
       q_{q}^\mu &=&
      \left(yzP^+, \ yz\vec P_{\!\perp} + z\vec q_{\!\perp}
      + \vec l_{\!\perp}, \ p_{g_1}^-\right)
\ ,  \nonumber \\
       \bar q_{\bar q}^\mu &=&
      \left(y(1-z)P^+, \ y(1-z)\vec P_{\!\perp} + (1-z)\vec q_{\!\perp}
      - \vec l_{\!\perp}, \ p_{g_2}^-\right)
\,,  \end {eqnarray}
we calculate first the propagators in the $q\bar q\,q\bar q$
and $q\bar q\,gg$ sectors,  
\begin  {eqnarray}
     \overline G _{q\bar q\,q\bar q} = \overline G _4 
      &=& -{q_g^+\over P^+}{z(1-z)\over z(1-z)Q^2 
      + \vec l_{\!\perp}^{\,2}+m^2_f }
\,,\quad{\rm and} \\  
      \overline G _{q\bar q\,gg} = \overline G _6 
      &=& -{q_g^+\over P^+}{z(1-z)\over z(1-z)Q^2 
      +\vec l_{\!\perp}^{\,2}}
\ .  \end{eqnarray}
They depend on $Q^2$, the momentum transfer along the quark line. 
The vacuum polarization diagrams are then calculable according
to the rules of light-cone perturbation theory \cite{bpp97} and yield
\begin{eqnarray}
      \Gamma_4 &=& - \sum_f {\alpha\over 4\pi} \int \!dz\,dl_{\perp}^2
      \ \Theta_s\ {1-2z(1-z)-{2m_f^2\over Q^2}
      \over l_{\perp}^2+m_f^2+z(1-z)Q^2 }
\,,\\ 
      \Gamma_5 &=&  \phantom{-\sum_f }{\alpha n_c \over 2\pi} 
      \int\! dz\,dl_{\perp}^2 \ \Theta_g\ {2-z(1-z)
      -{1\over z(1-z)}
      \over l_{\perp}^2+z(1-z)Q^2 }
\,.\end{eqnarray}
The dependence on $Q^2$ is caused by  the propagators.
The term with $1/z(1-z)$ in $\Gamma_5$
cancels with $\Gamma_1+\Gamma_2+\Gamma_3$, like in  
the previous work \cite{tho79,phz93}. These authors 
have however not kept track of the $Q^2$-dependence.  
Dropping this term replaces $\Gamma_5$ with
$\overline\Gamma_5$. 
The cut-off functions $\Theta$ are defined by
vertex regularization in Eq.(\ref{eq:i1}) and (\ref{eq:i2})
\begin {equation} 
       \Theta(z,l_{\!\perp}^2) = 
       \cases{
       1, &for $\ 0 \leq l_{\!\perp}^2 \leq  l^2_{\Lambda} (z)$
            and $\ \epsilon_l\leq z \leq 1 - \epsilon_u\ $ , \cr
       0, &otherwise.\cr}
\label{eq:i3}\end{equation}
The limiting transverse momentum $l_\Lambda^2(z)$ 
describes a semi-circle in the appropriate units, 
\begin{eqnarray} 
      l_\Lambda^2(z)&=&\big(\Lambda^2+(m_1+m_2)^2\big)
      \bigg[\bigg({c\over2}\bigg)^2 -\bigg(z-{b\over2}\bigg)^2\bigg]
\nonumber\\ \quad{\rm with}\quad
       b&=&{\Lambda^2+2m_1(m_1+m_2)\over
       \Lambda^2+(m_1+m_2)^2} \quad{\rm and}\quad
       c^2={\Lambda^2\over\Lambda^2+(m_1+m_2)^2}
       {\Lambda^2+4m_1m_2\over\Lambda^2+(m_1+m_2)^2}
\,.\end{eqnarray} 
The semi-circle is limited by its intersections at $z_1=\epsilon_l$
and $z_2=1 - \epsilon_u$, with 
\begin {equation} 
       \epsilon_l={b-c\over2} \quad{\rm and}\quad
       \epsilon_u=1-{b+c\over2} 
\,.\end{equation}
For sufficiently large $\Lambda$
and equal masses $m_1=m_2=m$, the limits become
$\epsilon_l=\epsilon_u= (m/\Lambda)^2$.
For the purpose of regularization, the gluons are endorsed 
with a small mass $m_g$, and thus vertex regularization 
regulates all divergences on the light cone. 
One gets straightforwardly
\begin{equation}
      \Gamma_4 = - \sum_f {\alpha\over 4\pi} 
      \int\limits_{\epsilon_f} ^{1-\epsilon_f} \!\!\!dz\,
      \bigg(1-2z(1-z)-{2m_f^2\over Q^2}\bigg)
      \ \ln{ {\Lambda^2+4m_f^2+Q^2
      \over Q^2+{m_f^2\over z(1-z)} } }
\,.\end{equation}
This is approximated by replacing the term $m_f^2/z(1-z)$
by its maximum at $z={1\over2}$, that is by $4m_f^2$.
The same type of approximation is done for $\Gamma_5$,
yielding finally
\begin{equation}
      \Gamma_4 =\sum_f {\alpha\over 6\pi} 
      \bigg({3m_f^2\over Q^2} -1\bigg)
      \ln{\bigg( 1+{\Lambda^2\over 4m_f^2+Q^2}\bigg)}
\,,\quad{\rm and}\quad
      \overline\Gamma_5 ={11\alpha n_c \over 12\pi} 
      \ln{\bigg(1+ {\Lambda^2\over 4m_g^2+Q^2}\bigg)}
\,.\label{eq:i62}\end{equation}
In either case, the terms depending on 
$\epsilon_f=(m_f/\Lambda)^2$ or 
$\epsilon_g=(m_g/\Lambda)^2$ 
have been dropped since they are small.
The effective coupling constant 
$\overline \alpha_s\equiv\alpha R^2_3=\alpha(1+\Gamma_4
      +\overline\Gamma_5)$, 
\begin{equation}
      \overline \alpha_s(Q^2) 
      = \alpha + {11\alpha^2 n_c \over 12\pi} 
      \ln{\bigg(1+{\Lambda^2\over 4m_g^2+Q^2}\bigg)}
      +{\alpha^2\over 6\pi} \sum_f 
      \bigg({3m_f^2\over Q^2} -1\bigg)
      \ln{\bigg(1+{\Lambda^2\over 4m_f^2+Q^2}\bigg)}
      \,,\label{eq:i63}\end{equation}
depends therefore on the momentum transfer.
Finally, the effective quark and gluon masses become
\begin{equation}
      \overline m_f^2 =m_f^2 +m_f^2 {\alpha\over \pi}
      {n_c^2-1\over 2n_c}
      \ln{ {\Lambda^2\over m_g^2} }
\,,\quad{\rm and}\quad
      \overline m_g^2 =m_g^2 -{\alpha\over 4\pi}
      \sum_f  m_f^2\ \ln{\bigg(1+{\Lambda^2\over 4m_f^2}\bigg) }
\,.\label{eq:i64}\end{equation}
They are obtained straightforwardly by light-cone perturbation 
theory \cite{bpp97} combined with vertex regularization.
The diagram for the effective quark mass 
$\overline m_f^2 =m_f^2(1+\Gamma_6)$
is given in Fig.~\ref{fig:g_3}. There is a corresponding
diagram for the effective gluon mass $\overline m_g^2$.

\section{Renormalization group analysis and confinement}
\label{sec:5}
Collecting terms and restricting to $n_c=3$ colors, the 
effective interaction as defined in  Eq.(\ref{eq:445})
becomes without the flavor changing interaction $U_a$
\begin{eqnarray} 
    E\ \langle x,\vec k_{\!\perp}; \lambda_{q},
    \lambda_{\bar q}  \vert \psi\rangle =\left[ 
    {\overline m^{\,2}_{q} + \vec k_{\!\perp}^{\,2}\over x } +
    {\overline m^{\,2}_{\bar  q} + \vec k_{\!\perp}^{\,2}\over 1-x }
    \right]\langle x,\vec k_{\!\perp}; \lambda_{q},
    \lambda_{\bar q}  \vert \psi\rangle &&
\nonumber\\
     - {1\over 3\pi^2}
    \sum _{ \lambda_q^\prime,\lambda_{\bar q}^\prime}
    \!\int_D\!dx^\prime d^2 \vec k_{\!\perp}^\prime\,
    \,{\overline\alpha_s(Q) \over Q  ^2} \,\langle
    \lambda_q,\lambda_{\bar q}\vert S(Q)\vert 
    \lambda_q^\prime,\lambda_{\bar q}^\prime\rangle
    \ \langle x^\prime,\vec k_{\!\perp}^\prime; 
    \lambda_q^\prime,\lambda_{\bar q}^\prime  
    \vert \psi\rangle &&
\,.\label{eq:i65}\end {eqnarray}
One recalls that 
the four-momentum transfer along the quark 
(or anti-quark) line is
$Q^2=-(k_q-k_q^\prime)^2=-(k_{\bar q}-k_{\bar q}^\prime)^2
\simeq (m_q +m_{\bar q})^2(x-x^\prime)^2 
+ (\vec k_{\!\perp}-\vec k_{\!\perp}^\prime)^2$.
As we are supposed to do, we restore the instantaneous
interaction $W$ in the $q\bar q$-space when deriving the above 
result, thus $\overline U_{q\bar q}
= W_{q\bar q} + \overline V\,\overline G_{q\bar q\,g}\overline V$.
Both the instantaneous interaction $W$ and the vertex interaction
$\overline V\,\overline G_{q\bar q\,g}\overline V$ have a 
non-integrable singularity, {\it cf} \cite{bpp97}, which cancels 
each other such that only the integrable Coulomb singularity 
remains. The latter manifests itself in Eq.(\ref{eq:i65}) in the 
``Coulomb denominator'' $Q^{-2}$.
The spinor factor $S(Q)$ represents the familiar current-current 
coupling which describes all fine and hyperfine interactions
\begin{eqnarray} 
    \,\langle
    \lambda_{q},\lambda_{\bar q}\vert S(Q)\vert 
    \lambda_{q}^\prime,\lambda_{\bar q}^\prime\rangle\,
    ={\left[ \overline u (k_q,\lambda_q) \,\gamma^\mu\,
    u(k_q^\prime,\lambda_q^\prime)\right] 
    \over \sqrt{x x^\prime} } \,
    {\left[ \overline u (k_{\bar q},\lambda_{\bar q}) 
    \,\gamma_\mu\,
    u(k_{\bar q}^\prime,\lambda_{\bar q}^\prime)\right] 
    \over \sqrt{(1 - x) (1- x ^\prime)} } 
\,.\end{eqnarray} 
One can simplify the expression technically by suppressing
in the latter  all momentum dependence. Evaluating the spinors 
at equilibrium, {\it i.e.}  at vanishing transverse momenta and at 
momentum fractions $\bar x_q = m_q /(m_q +m_{\bar q} )$ and
$\bar x_{\bar q}=m_{\bar q}/(m_q+m_{\bar q})$, 
gives \cite{bpp97}
\begin{eqnarray}
    {\left[ \overline u(k_q,\lambda_q)
    \,\gamma^\mu\,u(k_q^\prime,\lambda_q^\prime)\right] 
    \over \sqrt{x_{q}x_{q}^\prime}}
    {\left[\overline u(k_{\bar q},\lambda_{\bar q})\,\gamma_\mu\,
    u(k_{\bar q}^\prime,\lambda_{\bar q}^\prime)\right]\over
    \sqrt{x_{\bar q} x_{\bar q}^\prime}} \bigg\vert_{\rm equilibrium}
    = 4(m_q+m_{\bar q})^2\,
    \delta_{\lambda_{q},\lambda_{q}^\prime}
    \delta_{\lambda_{\bar q},\lambda_{\bar q}^\prime}
\,.\end{eqnarray}
Omitting the spin-dependence of the interaction implies to restrict
oneselve to the central part of the interaction. In this approximation
Eq.(\ref{eq:i65}) becomes
\begin{equation} 
    E\ \langle x,\vec k_{\!\perp}\vert \psi\rangle =\left[ 
    {\overline m^{\,2}_{q} + \vec k_{\!\perp}^{\,2}\over x } +
    {\overline m^{\,2}_{\bar  q} + \vec k_{\!\perp}^{\,2}\over 1-x }
    \right]\langle x,\vec k_{\!\perp}\vert \psi\rangle 
     - {4(m_q+m_{\bar q})^2\over 3\pi^2}
    \int_D\!\!\!dx^\prime d^2 \vec k_{\!\perp}^\prime
    \,{\overline\alpha_s(Q) \over Q  ^2} 
    \ \langle x^\prime,\vec k_{\!\perp}^\prime 
    \vert \psi\rangle 
\,.\label{eq:i68}\end{equation} 
It is this equation, finally, which we are going to analyze 
by the renormalization group.

\begin{figure} [t]
\begin{minipage}[t]{70mm} 
\begin{center}
\epsfxsize=60mm\epsfbox{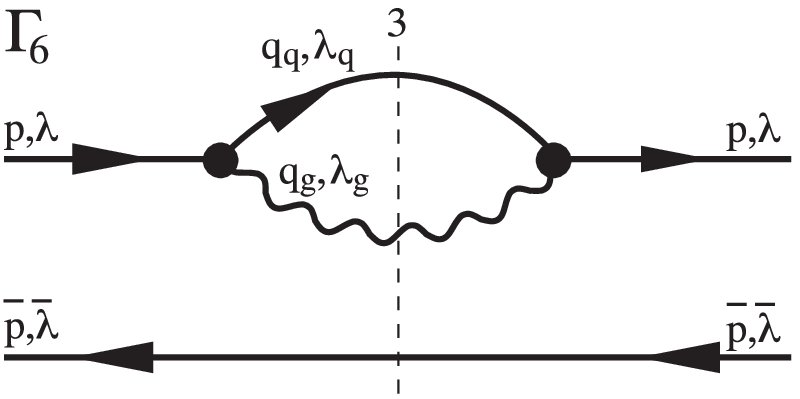}
\caption{\label{fig:g_3} 
The quark mass diagram.}
\end{center}
\end{minipage}
\hfill
\begin{minipage}[t]{70mm} 
\begin{center}
\begin {tabular}  {lr}
\hline
     $m_u=$ & 6 MeV \\ 
     $m_d=$ & 10 MeV \\ 
     $m_s=$ & 200 MeV \\ 
     $m_c=$ & 1.40 GeV \\ 
     $m_b=$ & 4.70 GeV \\ 
     $m_t=$ & 180 GeV \\ 
     $\alpha=$ & 0.2\qquad \\ \hline \hline
     $\Lambda=$ & 100 MeV \\ 
     $m_g=$ & 13 MeV \\ \hline
\end {tabular} 
\end{center}
\caption{\label{tab:quarks} 
Fixing the bare quantities.
}\vfill \end{minipage}
\end{figure}

In order to do calculations, one has to specify 10 ``input parameters''.
As specified in Fig.~\ref{tab:quarks}, these are the six flavor
masses $m_f$ \cite{pdg96}, the coupling constant 
$\alpha=g^2/4\pi$, and the two regularization parameters 
$\Lambda$ and $m_g$. To be concrete, these parameters 
are specified in Fig.~\ref{tab:quarks}, and subject to future change. 
The eigenvalues of the effective equation (\ref{eq:i68}) depend
on all of them, {\it i.e.} $E_i(m_f,\alpha;\Lambda,m_g)$.
While their dependence on the masses and coupling constant
is not unexpected, they should not depend on $\Lambda$, since
that is an arbitrary and unphysical parameter. One therefore 
must require that the eigenvalues satisfy 
the renormalization group equations
\begin{equation} 
    {d\over d\Lambda}\,E_i(m_f,\alpha;\Lambda,m_g) = 0
\,.\label{eq:i69}\end{equation}
Since the Lagrangian `bare' parameters themselves are arbitrary, 
one can try to let them be functions of $\Lambda$ in order to 
satisfy Eq.(\ref{eq:i69}): $m_f =m_f(\Lambda)$, 
$\alpha=\alpha(\Lambda)$, and $m_g=m_g(\Lambda)$. 
The numbers in Fig.~\ref{tab:quarks} can be viewed as such
functions at the particular value $\Lambda=100$ MeV.
If one finds one single set of functions which makes all eigenvalues 
$E_i$ independent of $\Lambda$, one has solved the problem.
It is not really necessary that $\Lambda\rightarrow\infty$.

In the present work, the eigenvalues of the full light-cone 
Hamiltonian $H_{LC}$ in Eq.(\ref{eq:2.2}) are conceptually 
identical with the eigenvalues of the effective Hamiltonian,
Eq.(\ref{eq:i68}). One must therefore analyze the latter. 
The dependence on the cut-off resides in three well separated 
locations: (1) in the domain of integration $D$, (2) in the effective
coupling constant $\overline \alpha_s(Q,\Lambda) $, and 
(3) in the effective mass parameters $\overline m_q(\Lambda)$
and $\overline m_{\bar  q}(\Lambda)$. As an advantage of vertex 
regularization, one can decouple the scales in $D$ and in the 
functions $\overline \alpha_s $ and $\overline m_q$. 
In the sequel we disregard the dependence on $D$.

If  $\overline \alpha_s $ and $\overline m_q$ do not dependent 
on $\Lambda$, the eigenvalues cannot depend on  $\Lambda$, 
either. It therefore suffices to require 
\begin{eqnarray}
    \delta\overline \alpha_s&=& 
    {\partial\overline \alpha_s\over\partial\alpha}\,\delta\alpha+ 
    {\partial\overline \alpha_s\over\partial\Lambda^2}
     \,\delta\Lambda^2+ \sum_f
    {\partial\overline \alpha_s\over\partial m_f^2}\,\delta m_f^2
     \equiv 0 
\,, \quad{\rm and} \label{eq:i70}\\
    \delta\overline m_{f'}^2&=& 
    {\partial\overline m_{f'}^2\over\partial\alpha}\,\delta\alpha+ 
    {\partial\overline m_{f'}^2\over\partial\Lambda^2}
     \,\delta\Lambda^2+ \sum_f
    {\partial\overline m_{f'}^2\over\partial m_f^2}\,\delta m_f^2 
     \equiv 0 
\,.\label{eq:i71}\end{eqnarray}
The variations of $\overline \alpha_s$ have to be 
performed of course at a fixed value of $Q^2$.
Solving the corresponding differential equations for 
$\alpha(\Lambda)$ and $m_f^2(\Lambda)$ yields the 
required functions. There are two regimes which are
particularly interesting. 
One is the {\em asymptotic regime}
\begin{equation}
    m_g^2,m_f^2 \ll Q^2 \ll \Lambda^2
\,,\end{equation}
in which Eqs.(\ref{eq:i62}) and (\ref{eq:i63}) become
\begin{equation}
    \overline \alpha_s= \alpha+\alpha^2 a_0\ln{\Lambda^2\over Q^2}
    \,,\qquad{\rm with}\quad a_0={33-2n_f\over 12\pi}
\,.\end{equation}
Solving Eq.(\ref{eq:i70}) in this regime yields 
{\em asymptotic freedom} in the front form, as shown first by 
Thorn \cite{tho79} and later by Perry {\it et al.} \cite{phz93}.
The asymptotic regime is particularly important for 
high energy scattering. But in a bound state calculation, 
the very large momentum transfers are less important than the
very small ones, because of  the Coulomb singularity 
$Q^2\sim0$, see Eqs.(\ref{eq:i65}) and (\ref{eq:i68}). 
It provides the binding. In the sequel we therefore emphasize 
the {\em bound state regime} near $Q^2\sim0$, 
\begin{equation}
    Q^2\ll m_l^2 \ll \Lambda^2\ll m_h^2
\,.\end{equation}
To our recollection, such a regime has not been analyzed 
by the renormalization group thus far. 
We shall differentiate between the two light quarks
with mass $m_l$ (u,d) and the four heavy quarks with mass 
$m_h$ (s,c,b,t). For convenience we collect some mass parameters
consistent with \cite{pdg96} in Fig.~\ref{tab:quarks}. In the present
context,  `light' and `heavy' is defined operationally by
\begin{equation}
    {m_l^2\over\Lambda^2}\ll 1\,,\quad{\rm and}\quad
    {m_h^2\over\Lambda^2}\gg 1
\,.\end{equation}
Consider first the effective gluon mass. In a gauge theory
the vector bosons should have zero mass 
$\overline m_g=0$. This in turn defines $m_g$
\begin{equation}
      m_g^2       ={\alpha\over 4\pi}\sum_h  
      m_h^2\ln{\bigg(1+{\Lambda^2\over 4m_h^2}\bigg)}
     +{\alpha\over 4\pi}
      \sum_l  m_l^2\ \ln{\bigg(1+{\Lambda^2\over 4m_l^2}\bigg)}
\,,\end{equation}
according to Eq.(\ref{eq:i64}). 
Expanding the logarithm for the former gives
\begin{equation}
      m_g^2={\alpha\over 4\pi}\Lambda^2\left(1+ \sum_l  
      {m_l^2\over\Lambda^2}\,
      \ln{\bigg(1+{\Lambda^2\over 4m_l^2}\bigg)}\right)
      \simeq {\alpha\over 4\pi}\Lambda^2
\,.\end{equation}
With the parameter values of Fig.~\ref{tab:quarks}, the numbers
in the parenthesis go like $(1+0.017)$. It is thus save to neglect 
the impact of the light quark masses. 
This yields $m_g\simeq 13$ MeV as quoted in the figure and 
removes the regulating gluon mass as a free parameter from 
the theory. 
In the vicinity of $Q^2\sim0$, the effective coupling constant 
goes like
\begin{eqnarray}
    \overline\alpha_s = {\alpha^2\over 2\pi Q^2} 
    \sum_f m_f^2 \ln{\bigg(1+{\Lambda^2\over 4m_f^2}\bigg)}
    &+&\alpha+{33\alpha^2\over 12\pi}  
    \ln{\bigg(1+{\Lambda^2\over 4m_g^2}\bigg)}
\nonumber\\
    &-&{\alpha^2\over 6\pi}  
    \sum_f \ln{\bigg(1+{\Lambda^2\over 4m_f^2}\bigg)}
    -{\alpha^2\over 8\pi}  
    \sum_f  {\Lambda^2\over \Lambda^2+4m_f^2}
\,.\end{eqnarray}
The last term is due to expanding the coefficient of $Q^2$
up to second order.
Treating differently the heavy and the light quarks again and 
substituting the gluon mass gives 
\begin{equation}
    \overline\alpha_s = {\alpha^2\Lambda^2\over 2\pi Q^2}  +
    \alpha + {33\alpha^2\over 12\pi} 
    \ln{\bigg(1+{\pi\over \alpha}\bigg)} - {\alpha^2\over8\pi} 
    \sum_l \bigg(1+{8\over3}\ln{{\Lambda\over 2m_l}}\bigg)
\,.\end{equation}
Near $Q^2\sim0$ the renormalization group equation 
for the effective coupling constant (\ref{eq:i70}) gives 
\begin{equation} 
    \delta\alpha\left[{\alpha\Lambda^2\over \pi Q^2}+1+\dots
    \right]+{\alpha^2\Lambda\over \pi Q^2}\,\delta\Lambda =0\,,
    \qquad{\rm thus}\quad \delta(\alpha\Lambda)=0
\,.\label{eq:i81}\end{equation}
Quite unexpectedly, $(\alpha\Lambda)$ turns out as a 
renormalization group invariant. The coupling constant 
``runs'' like 
\begin{equation} 
    \alpha(\Lambda)= {\alpha_0\Lambda_0\over\Lambda}
\,.\end{equation}
It is admitted to call this 
``asymptotic freedom in the bound state regime''.
The subscript $()_0$ denotes the renormalization point. 
A particular ``renormalization point'' is the set of numbers 
in Fig.~\ref{tab:quarks}. 
Finally, with the same approximations, the effective quark mass 
becomes according to Eq.(\ref{eq:i64}) 
\begin {equation} 
    \overline m_f^2 = m_f^2\bigg(1+{4\alpha\over3\pi}
    \ln{ {4\pi\over\alpha} }\bigg)
\,.\label{eq:i79}\end{equation}
Renormalizing the quark mass according 
to Eq.(\ref{eq:i71}) and introducing 
$t(\Lambda)=\ln{(4\pi/\alpha)}$ as an integration variable, thus
$t(\Lambda)=\ln{(4\pi\Lambda/[\alpha_0\Lambda_0])}$, 
gives 
\begin{equation} 
    \ln{{m_f(\Lambda)\over m_f(0)}}= 
    \int\limits^{t(\Lambda)}_{t(0)}\!\! dt
    \,{t-1\over t +{3\over 16} \exp{t} }
\,.\end{equation} 
One notes that the mass ratios $m_f/m_{f^\prime}
=\overline m_f/\overline m_{f^\prime}$
are renormalization group invariants, as they should.

In the bound state regime, the kernel of the integral equation 
(\ref{eq:i68}) becomes thus
\begin{equation}
    {\overline\alpha_s(Q)\over Q^2} = 
    {\kappa^2\over Q^4} + {\beta\over Q^2} +\dots
\,.\label{eq:i84}\end{equation}
Its Fourier transforms behave like $\beta/r + \kappa^2r$, 
{\it i.e.} like the superposition of a Coulomb and a
linear potential governed by a ``string constant $\kappa$'',
see for example \cite{pam95}.
The theory ``confines''. The two coefficients 
\begin{equation}
    \kappa = {\alpha_0\Lambda_0\over\sqrt{2\pi} }
    \,,\qquad{\rm and}\quad
    \beta=\alpha_0
    + {33\alpha_0^2\over 12\pi} 
    \ln{\bigg(1+{\pi\over \alpha_0}\bigg)} - {\alpha_0^2\over8\pi} 
    \sum_l \bigg(1+{8\over3}\ln{{\Lambda\over 2m_l}}\bigg)
\,,\label{eq:i85}\end{equation}
are related to each other by the renormalization group.
The string constant becomes a parameter of the theory
which must be determined by experiment. The value 
$\kappa\simeq10$ MeV consistent with Fig.~\ref{tab:quarks} 
is probably not very realistic.

Although the present approach has yet to be applied to hadrons,
the calculational scheme is rather different from the one 
taken by Wilson and collaborators \cite{wwh94}, see also
\cite{brp95,bpw97} and \cite{jop96a,jop96b}, mostly because
the present approach emphasizes so strongly the region
around $Q^2\sim 0$. A detailed comparison should be
given separately when dealing with renormalizing
the scales connected to the integration domain $D$ in 
Eqs. (\ref{eq:i65}) and (\ref{eq:i68}).
It is more than likely that the similarity transform of Wegner 
and collaborators \cite{weg94,lew96,guw97}
will be there more than useful.

In order to do calculations, one must give 8 parameters:
 the bare coupling constant, the renormalization scale 
and the six bare flavor masses $(\alpha,\Lambda,m_f)$.
As always in the sequel, we understand these parameters 
as given at the renormalization point and drop the subscript 
$()_0$. It is not relevant in this context whether one gives the
bare or the dressed masses, since they are related by 
the coupling constant in a simple way, Eq.(\ref{eq:i79}).
These are seven dimensionful and one dimensionless parameter.
We like to think of them in an other way. The relation between
$\Lambda$ and $\kappa$ is linear, see Eq.(\ref{eq:i85}).
Because of that one can introduce $(\alpha,\kappa,m_f)$ 
as independent variables and drop any explicit reference to 
$\Lambda$ as the parameter of regularization. 
Rather the string constant $\kappa$ sets the scale of the problem. 
The rest are dimensionless parameters, the coupling constant
or mass ratios, which are introduced to cope with the 
plenitude of experimental numbers. It looks as if  the
effective theory has one parameter more than the Lagrangian,  
but this is not quite true. In a Lagrangian theory, the absolute 
values of the spectral eigenvalues $E_i$ are physically 
meaningless. One can multiply the Lagrangian and thus 
the Hamiltonian with an arbitrary number without changing the
physics. In a way, renormalization theory fixes that number 
by explicitly introducing a scale.
 
Finally, one should emphasize that one is not completely free 
in choosing the scale:  The above formalism holds only for 
at least one `heavy' quark, with a mass much larger than 
$\kappa$. Although there is a long way 
until one reaches the 180 GeV of the top, one may ask 
what happens to the above when one transgresses 
this limit. If one repeats the above analysis for $Q^2\sim0$
but in the regime $Q^2 \ll m_f^2 \ll \Lambda^2$
the formalism changes.
The situation is similar to the asymptotic regime:
The coefficient of $Q^{-4}$ is roughly $m_f^2\alpha^2$,
thus unstable  with respect to the renormalization group.
As a consequence it fades away.
The result of a renormalization group analysis depends on 
the scale at which it is performed. 
In conclusion, we are afraid to 
pretend that the spectrum of toponium will be 
that one of a de-confining system, with an ionization threshold
like positronium. 

\section {Summary and discussion} 
\label {sec:6}

The present work has four important ingredients:
(1) The method of iterated resolvents is applied to a gauge 
theory Hamiltonian in the front form;
(2) In the solution the full quark-pair-gluon propagators 
can be well approximated by free propagators with the quarks
carrying the constituent mass $m=M/2$;
(3) Vertex as opposed to Fock space regularization is more
efficient to regulate all divergences in the light-cone formalism;
(4) After a renormalization group analysis, 
the spectrum of the Hamiltonian is independent of the
two regularization parameters and in the confining regime
a new mass scale appears which must by determined 
by experiment.~--- After the dust has settled, one seems 
to get a glimpse of an idea how confinement could 
work in practice. 

One should emphasize that in this non-perturbative approach 
there is no Fock space truncation. 
One calculates Lorentz scalars, the eigenvalues
of the invariant mass-squared, in a frame independent way.
Even the kernel of the interaction $\overline \alpha (Q)/Q^2$
is expressed in terms of an invariant, 
$Q^2=-(p^\prime-p)^2$, which is the momentum 
transfer of the quark.

The present work adopts systematically the point of view 
that all properties of a Lagrangian gauge field theory are 
contained in the canonical front-form Hamiltonian, which is 
calculated in the  light-cone gauge without the zero modes.
Periodic boundary conditions allow one to construct explicitly 
a Hamiltonian matrix. After regularization, one is confronted 
with the diagonalization of a large but finite matrix.
The binding of the particles in a field theory is provided
by the virtual scattering into the higher Fock spaces and 
the Tamm-Dancoff approach accounts for that, however 
it accounts for that only in lowest order. 
For large values of the coupling one loses control. 
These difficulties are overcome in the
the present work after re-analyzing the theory 
of effective interactions. 
By inverting the hierarchy, it was possible 
to develop a new formalism, the method of iterated resolvents,
and using this method, one can map the large Hamiltonian matrix  
onto  a matrix which is smaller. 
Finally, the periodic boundary conditions are relaxed 
by going over to the continuum limit.
Having found the eigenfunctions of the effective interaction,
one can retrieve all other many-body amplitudes 
in a self-consistent  way.

We saw that the effective interaction has two parts:
The flavor conserving interaction $U$ and flavor changing 
interaction $U_a$. Their diagrammatic representations
look like second order diagrams of perturbation theory, but 
represent a re-summation of perturbative graphs to all orders. 
Particularly, $U$ bears great similarity with a 
one-gluon-exchange interaction. In analyzing this effective
interaction one uses the apparent self-similarity in a gauge theory.
Ultimately, this aspect allows the `breaking of the hierarchy':
In the solution, the particles propagate like free particles. 
All many-body aspects reside in the vertex coupling functions, 
in which the well known divergences of field theory appear 
typically as $\vec k_{\!\perp}^{\,2}/x$. 
This can become large either for large transverse momentum 
$k_{\!\perp}$or for small longitudinal momentum fraction $x$.
In order to remove these divergences, one 
needs {\em two regulator scales}. 
Regulating the theory on the level of the fundamental
interaction, we introduce vertex regularization associated
with a mass scale $\Lambda$ and a small kinematical gluon 
mass $m_g$. Insisting that the gauge bosons have vanishing
dynamical mass, relates $m_g$ to $\Lambda$ and removes it 
from the formalism.

The appearance of a new scale $\kappa$ provides the room
for an interesting speculation. One knows empirically
that the invariant mass square  of the light mesons like the pion 
is linear in the quark masses. For dimensional reasons one needs
a parameter of dimension mass to account for that. 
One believes that this parameter is the $q\bar q$-vacuum 
condensate $\langle\overline\Psi\Psi\rangle$, {\it i.e.}
$M_{\pi}^2\sim(m_q+m_{\bar q})\langle\overline\Psi\Psi\rangle$.
The vacuum condensate is extremely difficult to calculate and
usually is taken from experiment.
With the appearance of the additional scale $\kappa$, 
which can play that role, one can 
have $M_{\pi}\sim (m_q+m_{\bar q}) \kappa$. Also $\kappa$ has
to be fitted to experiment, but its role within the theory is
well defined.

\end{document}